# Power Plexus: A network based analysis

Malvika Singh
DA-IICT
Gandhinagar, India.
201401428@daiict.ac.in

Sneha Mandan
DA-IICT
Gandhinagar, India
201401422@daiict.ac.in

Smriti Sharma
DA-IICT
Gandhinagar, India
201401003@daiict.ac.in

*Abstract*—Power generation and distribution remains an important topic of discussion since the industrial revolution. As the system continues to grow, it needs to evolve both in infrastructure, robustness and its resilience to deal with failures. One such potential failure that we target in this work is the cascading failure. This avalanche effect propagates through the network and we study this propagation by Percolation Theory and implement some solutions for mitigation. We have extended the percolation theory as given in [1] for random nodes to targeted nodes having high load bearing which is eliminated from the network to study the cascade effect. We also implement mitigation strategy to improve the network performance.

## I. INTRODUCTION

Cascading failures have tremendous impact on power grid networks. When the failure of a few nodes triggers the failure of other nodes which in turn cause the failure of other large number of nodes, it results in complete failure of power system. While some of such failures are smaller in magnitude because their growth is checked, in other cases it causes avalanche mechanisms. This was evident in the power failure mishap of 10th August 1996 [2], [3], when a 1300 Mega Watts electrical line in Oregon had failed and a chain reaction started which culminated in loss of power to more than 4 million people in 11+ states. This is also suspected to be the reason behind the last major power failure in the United States on August 14, 2003. Moreover, the redistribution of the power after failure of certain nodes leads to congestion and bottlenecks in the network as has been in the case of Internet congestion collapse, first recorded officially in October 1986, when the speed of connection between Lawrence Berkeley Laboratory and the University of Berkeley, two spots separated by a distance of 200m suffered a decline by a factor of 100 [2]. There have been works regarding the mitigation of such blackouts by considering inter-dependent networks of communication and network topology [4]. Other works include studying actual electromagnetic constructions and applications that go in power generation and transmission. However, such detailed analysis are extremely difficult to scale to networks of sizes having thousands of nodes. In our analysis, we have followed two approaches (i) Network performance based on size of giant component (ii) Cascading effect due to failure of a single node and dynamic redistribution of flows on the network [5]. Lastly we talk about a mitigation strategy.

## II. DESCRIPTION OF THE DATA SET

We have taken the data-set of US Electricity department [6] in this study, with N=4941 nodes and K=6594 edges. The electric power grid is represented as an undirected graph, in which the N nodes are the substations (generators, distribution substations or transformers) and the K edges are the transmission lines. To each edge between nodes i and j is associated a number $e_{ij}$ in the range [0; 1] measuring how efficiently nodes i and j communicate through the direct connection. For instance $e_{ij} = 1$ means that the arc between i and j is perfectly working, while $e_{ij} = 0$ indicates that there is no direct connection between nodes i and j. The weight of each edge can be understood as the cost of power transmission and is taken to be inversely proportional to the efficiency of the edge. With each node i are associated the characteristics - load ($L_i$) and threshold capacity of node ($C_i$) [7]. In our case, the load of a node is the betweenness centrality. (Betweenness centrality of a node v is the sum of the fraction of all-pairs shortest paths that pass through v). The capacity of the node is taken proportional to the initial load (betweenness centrality).

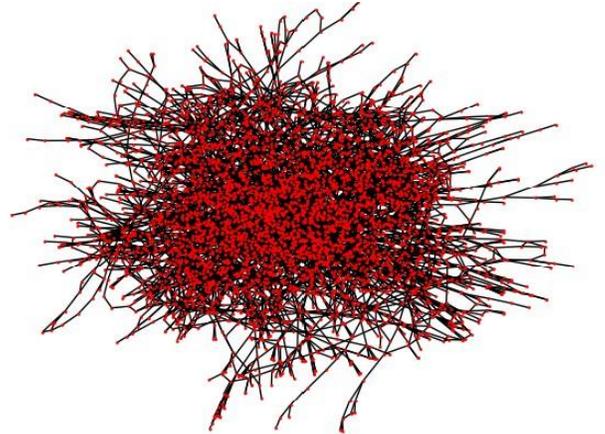

Fig. 1: Network of US power grid 2014

$$C_i = a * L_i, \ a \geq 1, i = 1, 2, .., N \quad (1)$$

where $C_i$ is the capacity of $i^{th}$ node, $L_i$ is the load of $i^{th}$ node and $a$ is the tolerance parameter

## III. MODEL

Percolation Theory as suggested in [1] refers to removal of nodes randomly from a network and study its effect on the remainder network. Here, in this work, we initially remove a node in two ways: 1. Randomly remove a node and 2. Remove node with highest betweenness centrality (modified percolation) and then after that the system is left to study the cascading effect on the remainder of the network as the threshold capacity of the remaining nodes exceeds its maximum capacity.

*A. Assumptions*

In this work, the load of a node is taken to be the measure of the metric - betweenness centrality. This is done because the load a node can carry is determined by the shortest paths to other nodes passing through that node. In actual physical system too, the voltage gets divided at the source depending upon the number of consumers that it has to supply electricity to. It is assumed that power transmission between the nodes takes the shortest path (path with minimum cost or the most efficient path).

Here, the cascading effect is triggered by initial node failure. All nodes are identical. Transformer, generator or substation are assumed to be similar.

Initially, all edges are considered to be identical, that is, they have similar weight and efficiency.

*B. Network Performance based on size of Giant Component*

Here, we remove a group of nodes and look at its impact on the network. The performance of the network is defined in terms of the size of the largest connected component.

Initially, due to failure of a group of nodes, certain nodes in the rest of the network get overloaded. If such congested nodes (node whose load exceeds its load carrying capacity) are also removed from the network, performance of the network decreases. We checked this for two different tolerance parameters (i.e, the ratio of carrying capacity and load).

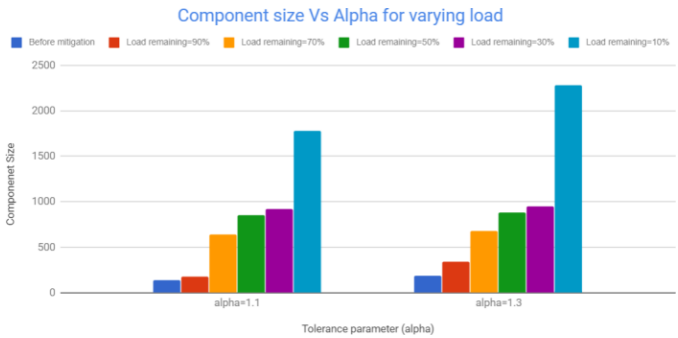

Fig. 2: Resultant component sizes during static analysis

*C. Cascading effect based on efficiency*

To simulate the cascading effect due to failure of an initial power station, we select an initial node that has failed and break its links with all its neighbours. The choice of node to be failed is done through two methods (i) Failure of random node (ii) Failure of node with very high load. After the initial failure of the node, the most efficient paths in the network change and hence the load of the nodes gets redistributed, eventually creating overload on certain nodes. For a congested node i , the efficiency of arcs at a time t is calculated by formula given in [7] and illustrated below between the node and its neighbours ($e_{ij}$) is reduced by the following rule, so that the power transmission takes an alternative route.

$$e_{ij}(t+1) = e_{ij}(0) * \frac{C_i}{L_i(t)} \qquad (2)$$

The efficiency of the network [8] is determined by the following equation:

$$E(tt) = \frac{1}{N(N-1)} \sum_{i,j \in G} s_{ij} = \frac{1}{N(N-1)} \sum_{i,j \in G, i \neq j} \frac{1}{d_{ij}} \qquad (3)$$

where $s_{ij}$ is the efficiency of the most efficient path between nodes i & j and $d_{ij}$ is shortest path between nodes i & j.

Thus, the overloaded nodes are not removed from the network. The efficiency of the power transmission through congested nodes decreases due to which such nodes will be avoided. The damage caused by the node failure is quantified by decrease in the efficiency E(G) of the network.

*D. Mitigation*

The mitigation strategy that we have simulated is based on Homogeneous Load Reduction [9]. The load of all nodes is reduced by certain percentage with an aim to limit the load below the threshold capacity of nodes after redistribution of loads. The number of congested nodes are considerably reduced after mitigation but some of them still remain congested after simulation.

The effect of mitigation in static analysis is seen in Fig.2. For $a = 1.1$, the number of congested nodes were 1274. On applying mitigation, the number of congested nodes dropped to 1108, 860, 632, 435 and 215 for remaining load fractions of 0.9, 0.7, 0.5, 0.3 and 0.1 respectively. For $a = 1.3$, the congested nodes previously were 1042 and after mitigation dropped to 922, 731, 541, 388, 197 for corresponding load fractions.

*E. Percolation*

The Percolation Concept used in this work deviates slightly from that proposed in [1]. It suggests a useful method to calculate robustness of the network in the form of Percolation Theory. It randomly eliminates a node from the network and studies the resultant effect. In this work, instead of randomly removing nodes, we initially remove that which carries higher load as compared to other nodes. After this step, we observe the cascading effect which changes the efficiency of the resultant network.

*1) Algorithm:* The time complexity of the algorithm is O(n*n*log(n)) as it calculates betweenness (all pair shortest path) and traverses for n nodes.

IV. RESULTS AND ANALYSIS

In this simulation we found out that by failing around 4 % of the total nodes in our data set the size of the giant connected component reduces to 57 %. This can be attributed to the percolation theory that we have used to calculate the

```
Algorithm 1 Cascading using Modified Percolation Algorithm
 1: procedure ALGORITHM(Graph g)
 2:     G ← edgelist(g)
 3:     weight_all_edges[] ← 1
 4:     init_efficiency ← efficiency(g)
 5:     alpha ← threshold_coefficient
 6:     for i ← 1 to total_num_nodes do
 7:         Load[i] ← betweenness_centrality_of_node(i)
 8:     end for
 9:     the_list ← highest_betweenness_centrality_nodes()
10:     for j ← 1 to total_num_nodes do
11:         threshold_capacity_node(j) ← betweenness_all_nodes[j] *alpha
12:     end for
13:     for k ← 1 num_nodes_to_be_failed do
14:         x ← random(the_list)
15:     end for
16:     RemoveNode(x)
17:     Update_Graph(g)
18:     Eficiency_after ← efficiency(g)
19: end procedure
```

Fig. 3: Cascading using Modified Percolation Algorithm

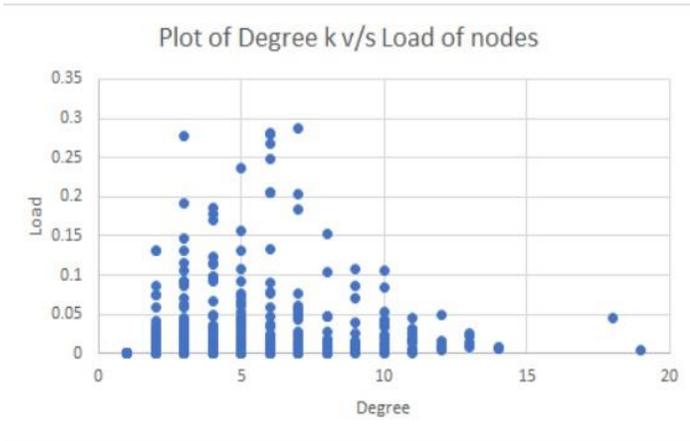

Fig. 4: Degree of nodes vs. the load

efficiency and the connected components.

We also observe that there is no correlation between the node of the degree and its load. So, the network topology is such that it is not necessary that a node having high degree has to bear the largest load (5).

The size of the biggest connected component is inversely proportional to the number of congested nodes. Mitigation causes a significant decrease in the number of congested nodes and consequently a bigger connected component. For $a = 1.3$, and an initial node removal of 50 nodes having highest load, mitigation by 90% load reduction was able to achieve a connected component covering 46% nodes vs. about 4% nodes without mitigation. From figure, the random removal of node does not have major impact on efficiency of the network whereas for tolerance parameter in the range $1 \leq a \leq 1.5$ for the load based node removal, the efficiency decreases notably with decrease in alpha. Also, the efficiency is for load based removal is considerably lesser than that for random removal

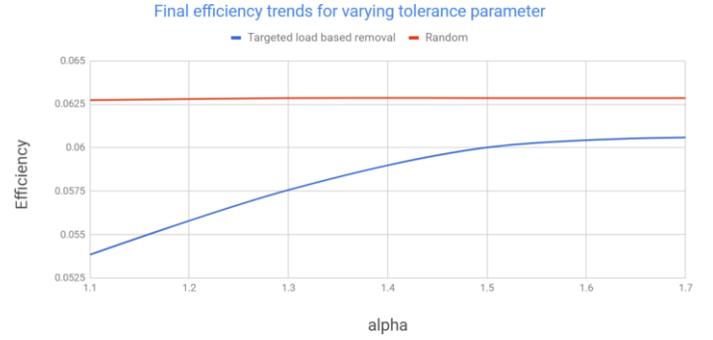

Fig. 5: Final efficiency for random and load based removal. (The efficiency is averaged over five iterations)

## V. CONCLUSION

In this work, we have studied the effect of cascading failure using site percolation theory in which we target the nodes bearing heavy load and the efficiency of the network before and after the failure is analyzed. We also study the resilience of the system in terms of targeted attacks or breakdown by calculating the size of operating parts after the cascading failure stops. It is found that upon breaking down of about 4% of the nodes due to failure, the size of the giant connected component is reduced to 57%. The mitigation strategy of reducing node load for the for network after the initial node failure was suggested. It is an easily implementable strategy and is reliable in improving network performance.

## VI. LIMITATIONS AND FUTURE WORK

The data set in consideration is large with N=4941 nodes and K=6594 edges and complex which required about 2 hours for computation for one combination of parameter values. This was the case when sequential execution was taken into account. The work can be extended to parallel execution using HPC Cluster for MPI and CUDA based parallelization of computationally expensive problems. The network can also be taken to be dynamically evolving in time by simulating addition of new nodes in place of failed nodes which will be deleted from the network. The attachment can happen guided by Albert Barabasi's random/preferential attachment and analyze both systems [10]. Also, edge percolation and its impact on the network can be studied. Different mitigation strategies such as Targeted range based load reduction can be implemented.